# Arterial stiffening provides sufficient explanation for primary hypertension

Klas H. Pettersen[1], Scott M. Bugenhagen[2], Javaid Nauman[3], Daniel A. Beard[2] & Stig W. Omholt[3]

[1]Department of Mathematical and Technological Sciences, Norwegian University of Life Science, Norway

[2]Department of Physiology, Medical College of Wisconsin, Milwaukee, Wisconsin, USA

[3]NTNU Norwegian University of Science and Technology, Department of Circulation and Medical Imaging, Cardiac Exercise Research Group, Trondheim, Norway

Correspondence should be addressed to: KHP (klas.pettersen@gmail.com)

**Author contributions:** K.H.P. and S.W.O. designed the study. K.H.P. constructed the integrated model and performed the numerical experiments with contributions from D.A.B. and S.M.B.. J.N. extracted and compiled empirical test data from the HUNT2 Survey. S.W.O, K.H.P. and D.A.B. wrote the paper.

## Abstract

**Hypertension is one of the most common age-related chronic diseases and by predisposing individuals for heart failure, stroke and kidney disease, it is a major source of morbidity and mortality. Its etiology remains enigmatic despite intense research efforts over many decades. By use of empirically well-constrained computer models describing the coupled function of the baroreceptor reflex and mechanics of the circulatory system, we demonstrate quantitatively that arterial stiffening seems sufficient to explain age-related emergence of hypertension. Specifically, the empirically observed chronic changes in pulse pressure with age, and the impaired capacity of hypertensive individuals to regulate short-term changes in blood pressure, arise as emergent properties of the integrated system. Results are consistent with available experimental data from chemical and surgical manipulation of the cardio-vascular system. In contrast to widely held opinions, the results suggest that primary hypertension can be attributed to a mechanogenic etiology without challenging current conceptions of renal and sympathetic nervous system function. The results support the view that a major target for treating chronic hypertension in the elderly is the reestablishment of a proper baroreflex response.**

## Introduction

The progressive increase in blood pressure with age is characterized by a greater increase in systolic blood pressure than diastolic blood pressure from the middle adult years (1). While systolic blood pressure continues to rise until the eighth or ninth decade, diastolic blood pressure tends to remain constant or decline after the fifth or sixth decade, leading to an accelerated rise in pulse pressure after age 50 years (2-4)**.** This rise in pulse pressure with advancing age is consistent with an increase in large artery stiffness (5) leading to a larger forward pressure wave (3). The pressing question we address here is why the autonomic nervous system, which controls blood pressure through modulating vascular resistance and cardiac output (6, 7), does not compensate for the increase in pulsatile load following stiffening of the arterial wall.

Because the arterial baroreceptors do not respond to pressure, but to strain (6), we hypothesized that the stiffening of the arterial wall (8, 9) may lead to constitutively reduced signaling from the baroreceptors to the barosensitive sympathetic efferents (6) at high pulse pressure. By misinforming the autonomic nervous system about the actual blood pressure and thus preventing it from exerting a proper negative feedback response through regulation of the heart rate, vasculature and renal system, the compromised baroreceptor function then hypothetically leads to an increasing baseline pulse pressure with increasing stiffening of the aortic wall.

The above hypothesis does not challenge currently accepted mechanisms for blood pressure regulation by the renal system (10). But it implies that the increase in sympathetic tone associated with increasingly more dysfunctional baroreceptor signaling with age shifts the renal pressure-diuresis/natriuresis function curve to higher pressures, because information about actual blood pressure to the renal system is conveyed through the sympathetic system based on baroreceptor response to strain.

Several authors have pointed to the possible etiological importance of reduced aortic compliance with age in relation to hypertension, but whether it plays a role in most cases of primary hypertension still remains unclear. To demonstrate the viability of our hypothesis in quantitative terms we integrated age-dependent arterial stiffening into a composite circulatory and baroreflex model in which the sympathetic and parasympathetic nervous activity regulate the heart rate in response to changes in blood pressure (Fig. 1A,B).

The lumped parameter model (see Methods for a detailed description) is based on the circulatory model by Smith et al. (11, 12), the baroreflex model of Bugenhagen et al. (13, 14), a model of the age-dependent aortic strain-pressure relationship established by Allen L. King in 1946 (15) (Fig. 1C), and a baroreceptor stimulus-response model following from standard receptive field theory of neurons (16). Guided by experimental data (17) we included adaptation of the baroreceptors through changes in the baroreceptor thresholds for the various age groups. By assuming a constant blood volume for all age groups, the model explicitly does not account for the regulation of plasma volume and salt through the kidney and the renin-angiotensin system (10) following from the hypothesized shift in the renal pressure-diuresis/natriuresis function curve. This allowed us to computationally probe whether the direct effects of stiffening on the system illustrated in Fig.1 are alone sufficient to explain the emergence of primary hypertension with age.

## Results and Discussion

The consequences of arterial stiffening on cardiovascular function were simulated for various age groups based on the aortic strain-pressure relationship that follows from increased stiffening of the arterial wall (15, 18) (Fig. 1C), while varying the total peripheral resistance (which is primarily determined by the contractile state of small arteries and arterioles throughout the body (19)) from 50% to 150% of baseline level. Using the predicted stroke volumes and heart rates across this peripheral resistance range, and assuming that a young individual has a mean cardiac output of about 5.5 L/min at baseline (20), and that this figure according to empirical data falls with about 0.25 L/min per decade (21, 22), enabled us to predict an approximately 1.75x linear increase of peripheral resistance across the focal age range (Fig 2C). This predicted relationship was then used to constrain the baseline peripheral resistance when calculating the central hemodynamic characteristics for the various age groups. Without changing any heart-specific parameters, the model predicts an almost linear decrease in stroke volume with age. Its predictions concerning temporal development of mean diastolic pressure and systolic pressure are concordant with available empirical data from the Framingham study (5) and the Norwegian HUNT2 Survey counting 62500 individuals (23) (Fig. 2A, B, D). While the diastolic pressure for all individuals is predicted to be categorized as normotensive or slightly prehypertensive according to the JNC-7 classification (24) (Fig. 2A (rectangles)), the systolic pressures of the three oldest age groups are predicted to be in the stage 1 hypertensive and severely stage 2 hypertensive groups.

While the trend of increasing systolic pressure with age emerges from the simulations, model predictions overestimate values of systolic pressure, in particular for the oldest group (71-78 yr) (1-4). Assuming fully functional baroreceptors, our model predicts that there is still almost no baroreflex signaling with changing blood pressure for this age group. Assuming this stands up to test, we conjecture that this almost total loss of excitatory input from baroreceptor afferents may be associated with reorganization of the neural activity within the nucleus of the solitary tract (NTS) (19) and perhaps other nuclei in the baroreflex as observed following sinoaortic denervation (SAD) (25-28); leading to a more inhibitory sympathetic tone. Since the model does not include such a mechanism, it does not account for the possibility that such an increase in inhibitory sympathetic tone may cause the kidneys to compensate through mechanisms of pressure-natriuresis-diuresis and the renin-angiotensin system. These mechanisms, which can reduce heart rate, increase renal blood flow and glomerular filtration rate, reduce renal sodium retention due to reduced aldosterone production, and reduce vascular tone (6, 19, 28), may explain the observed systolic pressure among elderly with autonomous nervous systems experiencing strong to almost total loss of excitatory input from the baroreceptor afferents.

The model predictions are dependent on the stipulated relationship between arterial distensibility and baroreflex signaling (i.e. baroreflex sensitivity (BRS)), obtained from combining the age-dependent aortic volume-pressure relationship developed by King (15) with our baroreceptor stimulus-response model based on standard receptive field theory of neurons (16). We tested the predicted relationship between BRS and age by first mimicking a standard Valsalva maneuver (i.e. inducing a brief temporal increase in thoracic pressure) on a young individual. Confirming that the model was indeed capable of predicting major features of the Valsalva maneuver in a young normotensive individual (Fig. 3A, B, C), we then used the *in silico* Valsalva maneuver to extract the BRS values for all age groups. The model results agree nicely with experimental data showing that cardiovagal baroreflex sensitivity declines progressively with age and is positively related to carotid artery compliance (7, 29) (Fig. 3D). Using these experimental data as test data instead of calibration data enabled us to make an independent assessment of a critical underlying premise of the integrated model.

According to Guyton's model of blood pressure regulation (10), any long-lasting alteration in blood pressure requires a shift of the kidney's acute pressure–natriuresis relationship (PNR). While this standard model acknowledges that renal dysfunction need not be the primary event in the cascade of changes leading to hypertension, it is asserted that whatever is the primary cause, it must lead to a change in the kidney's ability to excrete salt and water at a given level of blood pressure (30). There is no conflict between this assertion and our results. But our analysis shows that we do not need to invoke any pathophysiological change in kidney function, or to include a specific model of the renal system, to explain the emergence of hypertension with age. On the other hand, it is well documented that chronic elevation of renal perfusion pressure results in renal injury (arteriolar wall thickening, glomerular sclerosis and tubular sclerosis and interstitial sclerosis) (31). Thus, as renal function is diminished by pressure-induced injury, our model is fully concordant with the possibility that this may over time cause further elevation of blood pressure. Furthermore, it has been shown that renal dysfunction observed in the Dahl S rat model of hypertension may be explained as resulting primarily from stiffening of renal arterioles (32), consistent with the overall hypothesis explored here.

The mechanogenic hypothesis is intimately related to the fact that the baroreceptors do not respond to changes in blood pressure, but to changes in strain, and thus are likely to misinform the sympathetic system about the actual state of affairs when located in less compliant vessels. Recent experiments (33), demonstrating that the generally observed drop in blood pressure that follows from chronic stimulation of the carotid baroreflex can partly be attributed to sustained inhibition of renal sympathetic nerve activity, strongly supports this interpretation. It is also supported by data from renal denervation experiments (34, 35), which suggest that the sympathetic regulation of the kidneys, whether it is correctly informed or misinformed about the actual blood pressure by the baroreceptors, prevents activation of alternative regulatory mechanisms that apparently become invoked after denervation. It seems likely that these mechanisms at least in part cause a reduction in peripheral vascular resistance and/or blood volume (6, 36), two of the parameters in our model. Assuming normal renal function, the model predicts that even a moderate reduction in these parameters, due to diminished influence from a misinforming sympathetic control regime, will lead to the experimentally observed drop in pulse pressure of about 20 mmHg six months post treatment (34, 35) (Fig. 4).

A major argument against baroreceptor participation in determining blood pressure level is that they adapt to the prevailing pressure over time (37) and thus, cannot provide a sustained error signal to reflex mechanisms controlling the sympathetic nervous system. Our model based on a mechanogenic hypothesis is entirely consistent with observations on the "resetting" of baroreflex sensitivity in primary hypertension (38), and provides at least partial explanation for the phenomenon. The interpretation revealed by our analysis is that resetting, caused at least in part by mechanical remodeling of arteries, represents a primary cause rather than a consequence of hypertension. This is strongly supported by a comprehensive recent study by Kaess et al. (39) concluding that vascular stiffness appears to be a precursor rather than the result of hypertension. Indeed, for the observed levels of arterial stiffening with ageing to not represent a primary cause of hypertension, then adaptation or resetting of the baroreflex would have to be associated with a drastic increase in their sensitivity in order to counteract the reduction in compliance.

Our model-based analysis allows us to probe the potential influence of the arterial stiffening and the baroreflex system in isolation from the influence of other regulatory mechanisms influencing sympathetic activity beyond the baroreflex (10). Thus this analysis reveals that this mechanism on its own may explain the emergence of hypertension with age. By not including any compensation mechanism through adaptive changes in the heart or the vasculature to increase in blood pressure, we consider the discrepancy between model predictions and empirical data in Fig. 2 to strengthen the case for our mechanogenic hypothesis. If the model had predicted a weaker relation between blood pressure increase and age than empirically observed, this would have made the predominance of a mechanogenic mechanism much less likely. On the other hand, the empirical data on age-related hypertension clearly imply that compensatory mechanisms are able to only partially ameliorate the effects of arterial stiffening on blood pressure.

A lumped parameter model is per definition a model that simplifies the description of the behavior of a spatially distributed physical system into a topology consisting of discrete entities that approximates the behavior of the distributed system under certain assumptions. Considering the concordance between model predictions and empirical data, the resolution level of our model appears appropriate for what we set out to test in this paper. However, 3-D fluid-solid interaction models informed by better regional data than what are currently available would provide further valuable insight on both the local changes in hemodynamic loads and wall strain. As our model suggests that the aortic wall strain and baroreceptor output are key factors in blood pressure regulation, it strongly motivates the generation of such data and models. In particular, this would most likely lead to a better understanding of the etiology of hypertension at the individual level.

Our analysis illustrates the clear need for accounting for the ageing phenotype in efforts to understand the etiology of complex diseases. As the baroreceptors respond to strain and not pressure, the blood pressure regulatory system becomes dysfunctional when aortic strain, due to age-related stiffening, is no longer a good proxy for aortic blood pressure. The lack of mechanisms that fully compensate for the increasing aortic stiffness with age can easily be explained by standard evolutionary theory of aging (40).

Finally, our results suggest that arterial stiffness represents a therapeutic target by which we may be able to exploit an otherwise intact machinery for integrated blood pressure regulation.

**Methods**

**Model overview**

Our model is a composite of the circulatory model of Smith et al. (11) and the baroreflex model of Bugenhagen et al. (13), with modified heart dynamics, a new receptive field model for the baroreceptor stimulus-response relationship (16) and the King model of the aorta dynamics based on the age-dependent and nonlinear volume-pressure relationship derived from basic physical principles of elastomers (15).

Parameter values in the baroreflex model were set to original values reported in Bugenhagen et al. (13) for all components representing processes of the central nervous system activity, the dynamics of norepinephrine and acetylcholine at the sinoatrial (SA) node of the heart, and the effects of these concentrations of heart rate. For the unmodified parts of the cardiovascular system the original parameter values for the Smith et al. model were used. Below we focus on the novel model elements. One should consult original references (11, 13, 15) for the parts of the model that were not modified.

**Heart and circulatory system**

The original Smith et al heart and circulatory model (11), which simulates cardiac pumping at a constant heart rate $H$, was modified to use a variable input heart rate driving function $H(t)$, which is determined by the baroreflex model. The function $H(t)$, which depends on model-simulated acetylcholine and norepinephrine concentrations (13), is a continuous function of time $t$ and thus in general varies within one heart cycle. The complexity of the heart model was reduced by removal of the septum compartment, but the mechanical interaction between the heart ventricles was maintained through the shared pericardiac volume.

The cardiac domain contractilities/elastances are assumed to vary in proportion to the heart rate,

$$E_{es,lv} = E_{es0,lv}(1 + (H(t) - H_0)/3H_0),$$   (1)

$$E_{es,rv} = E_{es0,rv}(1 + (H(t) - H_0)/3H_0),$$   (2)

where a $30\%$ decrease in heart rate gives a $10\%$ decrease in elastance (41). The subscripts 'lv' and 'rv' denote the left ventricle and right ventricle, respectively. A constant value of $H(t) = H_0 = 80$ beats/minute gives the default elastance values $E_{es0}$ (11).

The linear pressure-volume relationships used in Smith et al. (11) are independent of the total blood volume and the model thus considers only the *stressed blood volumes* (with a total stressed blood volume of $1500 \, \mathrm{mL}$). Here, we used the non-linear pressure-volume relationship from King (15), and we assumed a total blood volume of 5000 mL. In the King model the pressure is given by a non-linear function of the relative volume $V_r = V/V_0 = (V_{ao} + V_{0,ao})/V_{0,ao}$,

$$P = A\left[V_r^{-1/3}\frac{\mathcal{L}^{-1}\{\beta V_r^{2/3}\}}{\mathcal{L}^{-1}(\beta)} - \frac{1}{V_r}\right], \qquad (3)$$

where $V_{ao}$ is the stressed aortic volume, $V_{0,ao}$ is the unstressed aortic volume, $\mathcal{L}$ is the Langevin function,

$$\mathcal{L}(z) = \coth(z) - \frac{1}{z}, \qquad (4)$$

and $\mathcal{L}^{-1}$ is the inverse Langevin function. The inverse Langevin function poses analytical challenges. However, within the domain of validity, $x \in (-1,1)$, the inverse Langevin function is well approximated with less than $5\%$ error at any point (42) by

$$\mathcal{L}^{-1}(x) \approx x\frac{3-x^2}{1-x^2}. \qquad (5)$$

We therefore made consequent use of this approximation.

In the King model the aorta is approximated as a cylinder: the aortic resting volume is given by $V_{0,ao} = \pi r_0^2 z_0$ and the stressed volume is given by $V = \pi r^2 z$, with $z_0$ and $r_0$ as the non-stressed and $z$ and $r$ as the stressed lengths and radiuses of the aortic cylinder, respectively. Further, the aortic wall is assumed to be perfectly elastomeric with the relationship between the length $z$ and radius $r$ given by $z = z_0(r_0/r)^{1/2}$. It then follows that the pressure can be expressed equivalently by the relative quantities $V_r$, $z/z_0$ or $r/r_0$. In the King model the unit reference volume ( $V_{0,ao} = 1\,\text{cm}$ ) is used, but in the present formalism reference volumes were chosen to give stressed volumes roughly in agreement with the original stressed volumes of the Smith model, which is about 140 mL at a pressure of 100 mmHg. This was achieved by setting $z_0 = 100\,\text{cm}$ for all ages. For an aortic pressure of 100 mmHg this gives total aortic volumes in the range from 248 mL (youngest) to 321 mL (oldest) and stressed aortic volumes in the range from 125 mL (oldest) to 142 mL (youngest) for the different ages.

**Baroreceptor afferent**

The relative volume $V_r$ is related to the aortic radius (15), $r$, through

$$V_r = \frac{V}{V_0} = (r/r_0)^{3/2} \, , \hspace{3cm} (6)$$

where $r_0$ is the non-stressed aortic radius. By using the definition of the strain,

$$\varepsilon = (r - r_0)/r_0 \, , \hspace{3cm} (7)$$

Eq. 3 gives the pressure-strain relationship

$$P = \frac{A}{\sqrt{\varepsilon + 1}} \left[ \frac{\mathcal{L}^{-1}\{\beta(\varepsilon + 1)\}}{\mathcal{L}^{-1}(\beta)} - \frac{1}{\varepsilon + 1} \right]. \hspace{1.5cm} (8)$$

In our model the strain is the input stimulus, to which the baroreceptor responds with a given firing rate. A linear stimulus-response model was constructed by expressing the linear firing rate $L$ as a convolution of the stimulus,

$$L(t) = \int_0^\infty D(\tau)\varepsilon(t - \tau)\mathrm{d}\tau \, , \hspace{2cm} (9)$$

where $D$ is the temporal kernel relating the stimulus to the response. A static nonlinearity function $F$ was introduced to model the firing-rate threshold. The non-linear firing rate, denoted $n$, can then be expressed as

$$n(t) = n_0 + F(L(t)) \, , \hspace{2.5cm} (10)$$

where $n_0$ is the background firing rate and $F$ is the linear threshold function (16),

$$F(L) = g[L - L_0]\theta(L - L_0) \, , \hspace{2cm} (11)$$

$\theta$ is the Heaviside step function, $L_0$ is the threshold value that $L$ must overcome to start firing, and $g$ is a proportionality constant. Figure 4A in Bugenhagen et al. (13), which is a reproduction of experimental results from Brown et al. (43), shows that experimentally induced steps in blood pressure give sharp overshoots of firing rate, followed by much slower saturations. Such an overshoot followed by a saturation can be modeled with a linear kernel $D(t)$ consisting of the two-exponential function,

$$D(t) = (\alpha e^{-t/\tau_1}/\tau_1 - e^{-t/\tau_2}/\tau_2)/(\alpha - 1) \, , \hspace{1cm} (12)$$

with time constants $\tau_2 > \tau_1$. The kernel $D$ is normalized so that the convolution integral, Eq. 9, gives $L = 1$ for stimulus $\varepsilon = 1$. The parameters $\alpha = 2.5$, $\tau_1 = 0.1\,\mathrm{s}$ and $\tau_2 = 0.5\,\mathrm{s}$ of Eq. 12 were found to give temporal responses to pressure step functions similar to the experiments.

In Andresen et al. (17) the gain $g$ and threshold $L_0$ are shown to express adaptation to increased stiffness of the aortic wall. In their Figure 5B two rats with different aortic stiffnesses are shown to express very different pressure-strain relationships, and their Figure 7C shows adaptation in the corresponding firing rate responses. The pressure-strain curves in their Figure 5B resemble the pressure-strain curves for ages 39 years and 75 years reported by King (15), and the corresponding firing rate threshold and gains, $L_0$ and $g$, were therefore used as thresholds and gains for the corresponding ages: $L_0(39 \text{ years}) = 87.5$, $L_0(75 \text{ years}) = 100$, $g(39 \text{ years}) = 0.52$ and $g(75 \text{ years}) = 0.32$. For the other ages the parameters for threshold and gain were intra- and extrapolated from these values.

The convolution formalism is tractable if the baroreflex is modeled as an open-loop process, which is not possible here as the baroreflex is part of a closed-loop system in which pressure influences the baroreflex afferent tone and the baroreflex efferent tone influences the pressure. Since the convolution kernel is expressed by decaying exponential functions, the convolution can, however, be transferred to equivalent differential equations (44). It can be shown (44) that the convolution integral given in Eq. 9 with an exponential kernel $D_a$,

$$D_a(\tau) = ae^{-a\tau}, \tau \geq 0, \qquad (13)$$

can be equivalently expressed by

$$\frac{\mathrm{d}L(t)}{\mathrm{d}t} = a[\varepsilon(t) - L(t)], \qquad (14)$$

with the initial conditions,

$$\frac{\mathrm{d}D_a(t)}{\mathrm{d}t} = -aD_a(t), D_a(0) = a. \qquad (15)$$

In our model the convolution integral can be split into two terms,

$$L(t) = (\alpha L_1(t) - L_2(t))/(\alpha - 1) \qquad (16)$$

with

$$L_1(t) = \int_0^\infty \frac{e^{-t/\tau_1}}{\tau_1} \varepsilon(t - \tau) \mathrm{d}\tau, \qquad (17)$$

and

$$L_2(t) = \int_0^\infty \frac{e^{-t/\tau_2}}{\tau_2} \varepsilon(t-\tau) \mathrm{d}\tau . \qquad (18)$$

The corresponding differential equations will then be

$$\frac{\mathrm{d}L_1}{\mathrm{d}t} = [\varepsilon(t) - L_1(t)]/\tau_1 , \qquad (19)$$

and

$$\frac{\mathrm{d}L_2}{\mathrm{d}t} = [\varepsilon(t) - L_2(t)]/\tau_2 . \qquad (20)$$

Thus, given the stimulus $\varepsilon(t)$, $L_1$ and $L_2$ are determined from Eqs. 19 and 20. Overall response $L$ is computed from Eq. 16 and the baroreceptor firing rate $n$ is given by Eq. 10.

## Acknowledgements

This work was supported by the Research Council of Norway under the eVITA program, project number 178901/V30, and by the Virtual Physiological Rat Project funded through NIH grant P50-GM094503. NOTUR, the Norwegian metacenter for computational science, provided computing resources under project nn4653k. We thank two anonymous reviewers for insightful comments and suggestions and the Nord-Trøndelag Health Study (The HUNT Study) for providing empirical data.

## References

1. Kannel WB, Gordan T (1978) Evaluation of cardiovascular risk in the elderly: the Framingham study. *Bull N Y Acad Med* 54:573–591.

2. Franklin SS, Khan SA, Wong ND, Larson MG, Levy D (1999) Is pulse pressure useful in predicting risk for coronary heart disease?: The Framingham Heart Study. *Circulation* 100:354–360.

3. Mitchell GF et al. (2010) Hemodynamic Correlates of Blood Pressure Across the Adult Age Spectrum: Noninvasive Evaluation in the Framingham Heart Study. *Circulation* 122:1379–1386.

4. Khattar RS, Swales JD, Dore C, Senior R, Lahiri A (2001) Effect of Aging on the Prognostic Significance of Ambulatory Systolic, Diastolic, and Pulse Pressure in Essential Hypertension. *Circulation* 104:783–789.

5. Franklin SS et al. (1997) Hemodynamic patterns of age-related changes in blood pressure: the Framingham Heart Study. *Circulation* 96:308.

6. Guyenet PG (2006) The sympathetic control of blood pressure. *Nat Rev Neurosci* 7:335–346.

7. Monahan KD (2007) Effect of aging on baroreflex function in humans. *Am J Physiol Regul Integr Comp Physiol* 293:R3–R12.

8. Zieman SJ (2005) Mechanisms, Pathophysiology, and Therapy of Arterial Stiffness. *Arterioscler Thromb Vasc Biol* 25:932–943.

9. McVeigh GE, Bank AJ, Cohn JN (2007) Arterial compliance. *Cardiovasc Med*:1811–1831.

10. Guyton AC (1991) Blood pressure control--special role of the kidneys and body fluids. *Science* 252:1813–1816.

11. Smith BW, Chase JG, Nokes RI, Shaw GM, Wake G (2004) Minimal haemodynamic system model including ventricular interaction and valve dynamics. *Med Eng Phys* 26:131–139.

12. Smith BW, Geoffrey Chase J, Shaw GM, Nokes RI (2005) Experimentally verified minimal cardiovascular system model for rapid diagnostic assistance. *Control Eng Pract* 13:1183–1193.

13. Bugenhagen SM, Cowley AW, Beard DA (2010) Identifying physiological origins of baroreflex dysfunction in salt-sensitive hypertension in the Dahl SS rat. *Physiol Genomics* 42:23–41.

14. Beard DA et al. (2012) Multiscale Modeling and Data Integration in the Virtual Physiological Rat Project. *Ann Biomed Eng*.

15. King AL (1946) Pressure-Volume Relation for Cylindrical Tubes with Elastomeric Walls: The Human Aorta. *J Appl Phys* 17:501.

16. Dayan P, Abbott LF (2001) *Theoretical Neuroscience: Computational and Mathematical Modeling of Neural Systems (Computational Neuroscience)* (The MIT Press). 1st Ed.

17. Andresen MC, Krauhs JM, Brown AM (1978) Relationship of aortic wall and baroreceptor properties during development in normotensive and spontaneously hypertensive rats. *Circ Res* 43:728–738.

18. Hallock P, Benson IC (1937) Studies on the elastic properties of human isolated aorta. *J Clin Invest* 16:595–602.

19. Coffman TM (2011) Under pressure: the search for the essential mechanisms of hypertension. *Nat Med* 17:1402–1409.

20. Proctor DN et al. (1998) Influence of age and gender on cardiac output-V O 2

relationships during submaximal cycle ergometry. *J Appl Physiol* 84:599–605.

21.  Fagard R, Thijs L, AMERY A (1993) Age and the Hemodynamic Response to Posture and Exercise. *Am J Geriatr Cardiol* 2:23–40.

22.  Stratton JR, Levy WC, Cerqueira MD, Schwartz RS, Abrass IB (1994) Cardiovascular responses to exercise. Effects of aging and exercise training in healthy men. *Circulation* 89:1648–1655.

23.  Holmen J et al. (2003) The Nord-Trøndelag Health Study 1995–97 (HUNT 2): objectives, contents, methods and participation. *Norsk epidemiologi* 13:19–32.

24.  Chobanian AV et al. (2003) The Seventh Report of the Joint National Committee on Prevention, Detection, Evaluation, and Treatment of High Blood Pressure: the JNC 7 report. *JAMA* 289:2560–2572.

25.  Cowley AW, LIARD JF, Guyton AC (1973) Role of the Baroreceptor Reflex in Daily Control of Arterial Blood Pressure and Other Variables in Dogs. *Circ Res* 32:564–576.

26.  Schreihofer AM, Sved AF (1992) Nucleus tractus solitarius and control of blood pressure in chronic sinoaortic denervated rats. *Am J Physiol* 263:R258–66.

27.  Ito S, Sved AF (1997) Influence of GABA in the nucleus of the solitary tract on blood pressure in baroreceptor-denervated rats. *Am J Physiol Regul Integr Comp Physiol* 273:R1657–R1662.

28.  Thrasher TN (2004) Baroreceptors, baroreceptor unloading, and the long-term control of blood pressure. *Am J Physiol Regul Integr Comp Physiol* 288:R819–R827.

29.  Monahan KD et al. (2001) Age-associated changes in cardiovagal baroreflex sensitivity are related to central arterial compliance. *Am J Physiol Heart Circ Physiol* 281:H284–H289.

30.  Malpas S (2009) Editorial comment: Montani versus Osborn exchange of views. *Experimental Physiology* 94:381–382.

31.  Mori T et al. (2008) High Perfusion Pressure Accelerates Renal Injury in Salt-Sensitive Hypertension. *Journal of the American Society of Nephrology* 19:1472–1482.

32.  Beard DA, Mescam M (2012) Mechanisms of pressure-diuresis and pressure-natriuresis in Dahl salt-resistant and Dahl salt-sensitive rats. *BMC Physiol* 12:6.

33.  Iliescu R, Irwin ED, Georgakopoulos D, Lohmeier TE (2012) Renal Responses to Chronic Suppression of Central Sympathetic Outflow. *Hypertension* 60:749–756.

34.  Krum H et al. (2009) Catheter-based renal sympathetic denervation for resistant hypertension: a multicentre safety and proof-of-principle cohort study. *Lancet* 373:1275–1281.

35. Mahfoud F et al. (2012) Renal Hemodynamics and Renal Function After Catheter-Based Renal Sympathetic Denervation in Patients With Resistant Hypertension. *Hypertension* 60:419–424.

36. Vink EE, Blankestijn PJ (2012) Evidence and Consequences of the Central Role of the Kidneys in the Pathophysiology of Sympathetic Hyperactivity. *Front Physio* 3.

37. Cowley A Jr (1992) Long-term control of arterial blood pressure. *Physiol Rev* 72:231–300.

38. Mancia G, Ludbrook J, Ferrari A, Gregorini L, Zanchetti A (1978) Baroreceptor reflexes in human hypertension. *Circ Res* 43:170–177.

39. Kaess BM et al. (2012) Aortic stiffness, blood pressure progression, and incident hypertension. *JAMA* 308:875–881.

40. Kirkwood TBL (1977) Evolution of ageing. *Nature* 270:301–304.

41. Nakayama Y et al. (2001) Heart Rate-Independent Vagal Effect on End-Systolic Elastance of the Canine Left Ventricle Under Various Levels of Sympathetic Tone. *Circulation* 104:2277–2279.

42. Cohen A (1991) A Padé approximant to the inverse Langevin function. *Rheologica Acta* 30:270–273.

43. Brown AM, Saum WR, Tuley FH (1976) A comparison of aortic baroreceptor discharge in normotensive and spontaneously hypertensive rats. *Circ Res* 39:488–496.

44. Smith H (2011) in *Texts in Applied Mathematics*, Texts in Applied Mathematics. (Springer New York, New York, NY), pp 119–130.

**Figures**

a

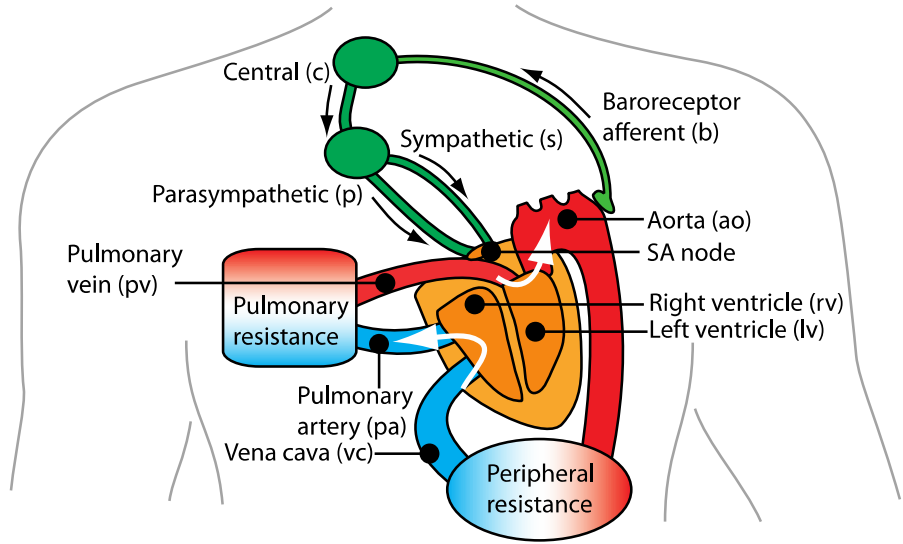

b

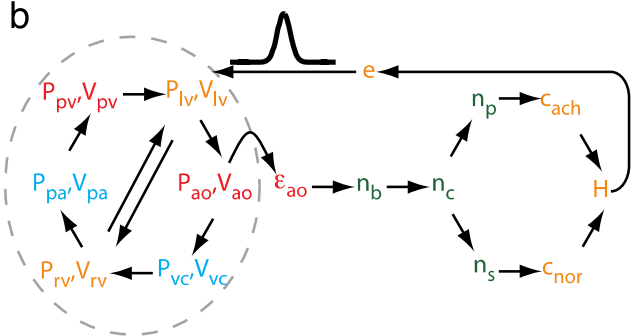

c

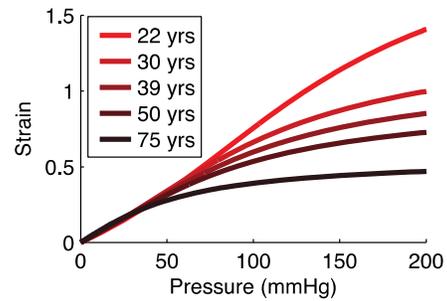

Figure 1. **Model overview**. (A) Schematic illustration of the anatomical structures contained in the model. (B) Model variables and their interconnections: $P$ is pressure; $V$ is volume; $\varepsilon_{ao}$ is aortic wall transversal strain; $n$ is firing rate; $c$ is concentration; $H$ is beat driver; $e$ is the beat driver function, which produces heart beats through the dynamic contribution to the pressure-volume relationships of the heart chambers. (C) Strain-pressure relationships for various age groups (15, 18) used in the integrated model.

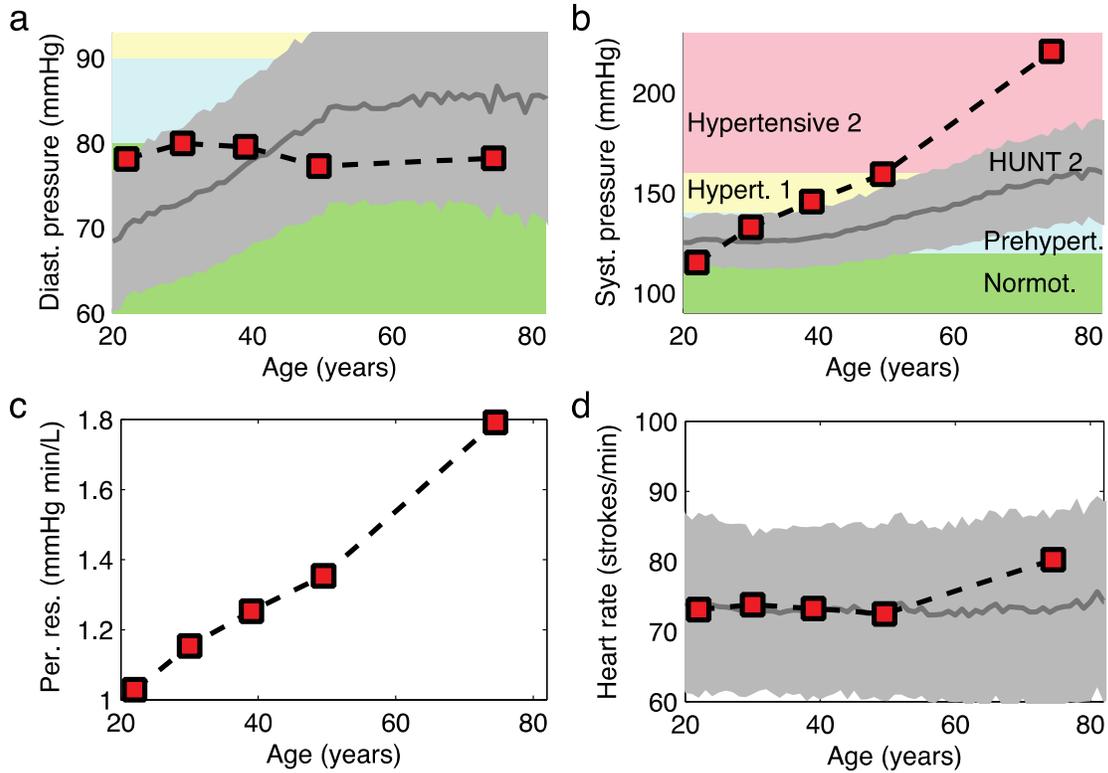

Figure 2. **Predicted age-dependent central hemodynamics**. Steady state values of (A) diastolic pressure, (B) systolic pressure, (C) peripheral resistance and (D) heart rate, obtained by assuming an age-dependent cardiac output (see main text). The squares refer to the age groups depicted in Fig. 1C. The solid grey lines show recorded mean values (± SD in grey) from 62496 individuals in the age range 20-80 years obtained from the Norwegian HUNT2 Survey (23). When assessing the fit between predicted and experimental data it should be emphasized that the HUNT2 data also include all individuals (8396) that were under antihypertensive therapy. The categorization from Normotensive (green) to Hypertensive 2 (red) is based on the JNC 7 classification (24).

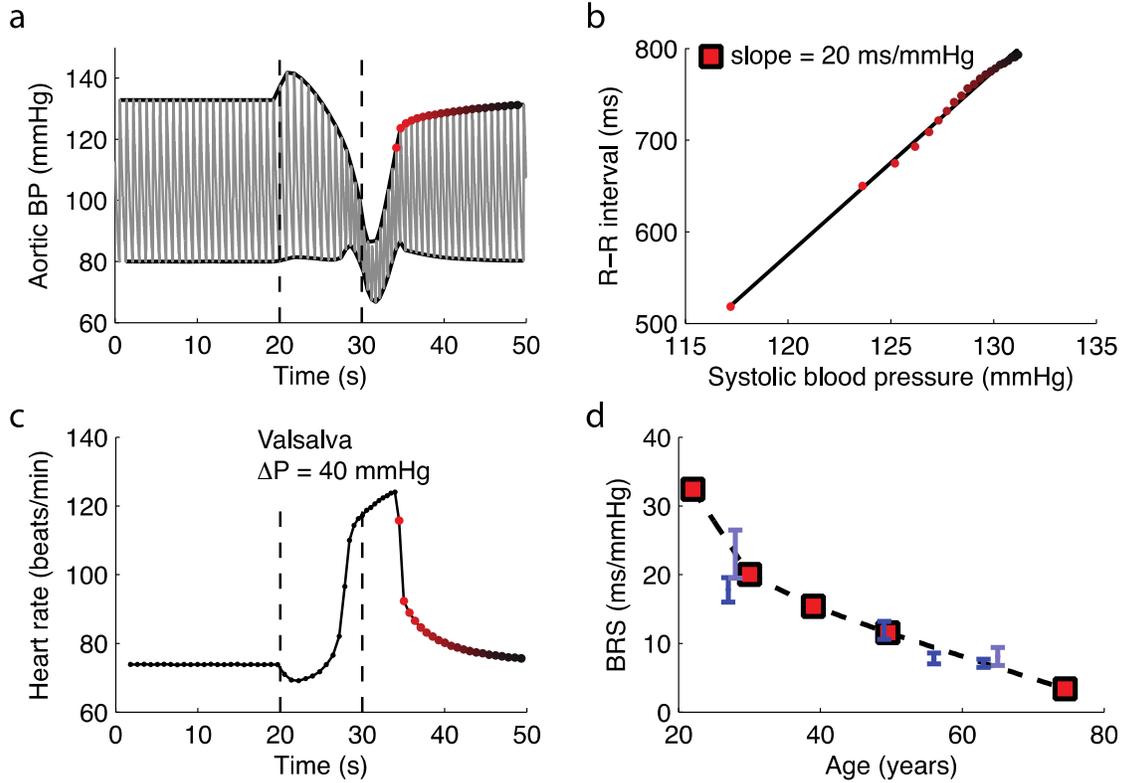

Figure 3. **Model response to Valsalva maneuver**. (A) Predicted changes in aortic blood pressure following from exposing individuals around 30 years to the Valsalva maneuver. The maneuver was mimicked by an increase in thoracic pressure of 40 mmHg, starting at t = 20 s and lasting for 10 seconds. (B) The baroreflex sensitivity (BRS) is computed by finding the slope of $\Delta$(R-R interval)/$\Delta$P after the heart rate in **c** has reached its peak, indicated by the corresponding red dots in A and C. Here, the R-R interval is given from the inverse of the heart rate in C. (D) Comparison between predicted BRS values for all ages and available experimental data (29).

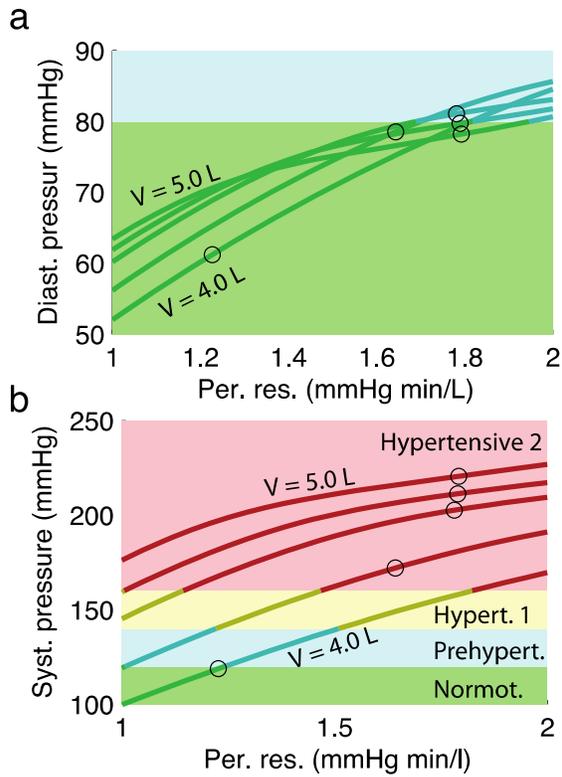

Figure 4. **Effect of reducing blood volume and peripheral resistance for the oldest age group (75 years).** The four curves show results for different blood volumes, and the colors indicate blood pressure categories (see Legend to Fig. 2). For both panels the five curves correspond to a blood volume of 5.0 L, 4.7 L, 4.5 L, 4.2 L and 4.0 L (starting from the top). (A) Diastolic aortic pressure. (B) Systolic aortic pressure.